\documentclass[aps,pra,twocolumn,showpacs]{revtex4}

\usepackage{amsmath}
\usepackage{amssymb}
\usepackage{bm}
\usepackage{epsfig}

\newcommand{\bb}{{\bf b}}
\newcommand{\bB}{{\bf B}}
\newcommand{\be}{{\bf e}}
\newcommand{\bg}{{\bf g}}
\newcommand{\bS}{{\bf S}}
\newcommand{\bU}{{\bf U}}
\newcommand{\bfeta}{{\bm \eta}}

\newcommand{\hUp}{\hat {\tilde U}}
\newcommand{\gp}{\tilde g}
\newcommand{\bp}{\tilde B}
\newcommand{\etap}{\tilde \eta}

\newcommand{\bgp}{\tilde{\bf g}}

\newcommand{\fp}{\tilde f}
\newcommand{\Cp}{\tilde C}
\newcommand{\LZ}{{\rm LZ}}

\newcommand{\tn}{\tau_{\rm n}}
\newcommand{\tacc}{\tau_{\rm acc}}

\newcommand{\ti}{{t_0}}
\newcommand{\cc}{\textrm{cc}}

\newcommand{\hbS}{\hat {\bf S}}
\newcommand{\hbsig}{\hat {\bm \sigma}}

\newcommand{\tx}{t_\times}

\newcommand{\figwidth}{8cm}

\begin{document}

\title{Correlations in noisy Landau-Zener transitions}

\author{V.L. Pokrovsky}

\affiliation{Department of Physics, Texas A\&M University, College
  Station, TX 77843-4242}

\affiliation{Landau Institute for Theoretical Physics,
Chernogolovka, Moscow District, 142432, Russia}

\author{S. Scheidl}

\affiliation{Institut f\"ur Theoretische Physik, Universit\"at zu
  K\"oln, Z\"ulpicher Str. 77, D-50937 K\"oln, Germany}

\date{\today}

\begin{abstract}
  We analyze the influence of classical Gaussian noise on Landau-Zener
  transitions during a two-level crossing in a time-dependent regular
  external field. Transition probabilities and coherence factors
  become random due to the noise. We calculate their two-time
  correlation functions, which describe the response of this two-level
  system to a weak external pulse signal. The spectrum and intensity
  of the magnetic response are derived.  Although fluctuations are of
  the same order of magnitude as averages, the results is obtained in
  an analytic form.
\end{abstract}


\maketitle

\section{Introduction}

The Landau-Zener (LZ) theory \cite{Landau,Zener} (see also
\cite{Stuckelberg}) plays an important role in many different physical
problems ranging from chemistry, biology, theory of collisions to the
tunneling of Bose-condensates, dynamics of glasses and spin reversal
in magnetic wires and nanomagnets, as well as quantum computing. LZ
theory treats a rather simple and general situation of two-level
crossing. Near the point of (avoided) level crossing, adiabaticity is
violated and transitions occur. The LZ Hamiltonian can be represented
by a $2 \times 2$ matrix acting in a two-dimensional space of vectors
spanned by the basis vectors $|{\uparrow}\rangle $ and
$|{\downarrow}\rangle$,
\begin{eqnarray}
  H_{LZ}=\left(
    \begin{array}{cc}
      E_{\uparrow} & \Delta \\
      \Delta ^{*} & E_{\downarrow}
    \end{array}
  \right) \ ,
  \label{H.LZ}
\end{eqnarray}
where $E_{\uparrow}=-E_{\downarrow}=-\frac{\hbar }{2}\nu t$ and the
transition matrix element $\Delta $ does not depend on time. Without
loss of generality one can assume $\Delta $ to be a positive constant.
Landau and Zener have found the evolution matrix
\begin{eqnarray}
  U_{LZ}=\left(
    \begin{array}{cc}
      a & b \\
      -b^{*} & a^{*}
    \end{array}
  \right)
\end{eqnarray}
which connects initial amplitudes in the diabatic basis \cite{note1}
with their final values.  The entries
\begin{eqnarray}
  a=e^{- \pi \gamma^2},
  &&
  b= - \frac{\sqrt{2\pi}}  {\gamma \Gamma(i\gamma^2)}
  e^{-\pi \gamma^2/2-i\pi/4}
\end{eqnarray}
depend only on the dimensionless parameter
\begin{eqnarray}
  \gamma =\frac{\Delta }{\hbar \sqrt{\nu }}
\end{eqnarray}
which we call the LZ parameter.

In many applications the LZ theory must be modified to take into
account thermal noise or noise of a different nature. In a pioneering
work Kayanuma \cite{Kayanuma} has solved such a problem on the basis
of three simplifying assumptions: i) the regular transition matrix
element is equal to zero (i.e., $\Delta =0$), the noise is completely
responsible for transitions; ii) the correlation function of noise has
a simple exponential form; iii) the noise is fast. The first two
limitations were lifted in the work by Pokrovsky and Sinitsyn
\cite{PS1}.  Their analysis of time scales for different processes
distinguishes the correlation time of noise $\tau _{ \mathrm{n}}$, the
accumulation time $\tau _{\mathrm{acc}}=(\nu \tau _{
  \mathrm{n}})^{-1}$ during which the noise effectively produces
transitions, and the LZ time $\tau _{LZ}=\Delta /(\hbar \nu )$ during
which the standard LZ transition (without noise) proceeds. The noise
is called fast if $\tau _{\mathrm{n}}\ll \tau _{LZ}\ll \tau
_{\mathrm{acc}}$. In this case the separation of times allows to
neglect the regular matrix element $\Delta $ at $|t|\gg \tau _{LZ}$
and to neglect the noise in the interval $|t|\ll \tau
_{\mathrm{acc}}$. Since these intervals overlap, one can solve first
the reduced problem with the noise producing transitions and then
match this solution with the LZ solution. In this way the authors of
Ref.~\cite{PS1} have found the average value of the density
matrix.

The two-level problem is as usual equivalent to the problem of
spin $\frac12$ rotating in the time-dependent field containing
regular and stochastic parts.  In their second work \cite{PS2} the
same authors represented the density matrix in terms of the Bloch
vector $\mathbf{g}$,
\begin{eqnarray}
  \hat{\rho}(t)=\frac{1}{2}[\hat{1}+\mathbf{g}(t)\cdot \hbsig] \ ,
\label{rho}
\end{eqnarray}
where $\hbsig$ is the triplet of Pauli matrices for spin
$\frac{1}{2}$. It is a well known fact that the square of the
Bloch vector is an integral of motion. In Ref.~\cite{PS2}
this fact was used to calculate the square fluctuation
\[
\langle [\delta \mathbf{g}(t)]^{2}\rangle =\mathbf{g}^{2}-\langle \mathbf{g}
(t)\rangle ^{2}.
\]
However, the fluctuations of separate components, $\langle [\delta
g_{i}(t)]^{2}\rangle $, ($i=x,y,z$) were not calculated, neither
two-time correlators $\langle g_{i}(t)g_{j}(t^{\prime })\rangle $ were
found.

Such fluctuations are observable, for example, in a system of magnetic
molecules subject to the same magnetic field (with identical regular
and random contributions).  The measurement of the $z$ component of
the magnetic moment averaged over the system yields $m_z(t) = \frac
\hbar2 \textrm{ Tr } \hat \sigma_z \cdot \hat \rho(t) = \frac \hbar2
g_z(t)$ which depends on the particular noise history of the
measurement.  The repetition of such a measurement for many
independent noise histories then yields an average $\langle m_z(t)
\rangle$ with fluctuations $\langle [\delta m_z]^2 \rangle =
\frac{\hbar^2}4 \langle [\delta g_z(t)] \rangle$.  Similarly, the
twofold time dependence of the correlators $\langle
g_{i}(t)g_{j}(t^{\prime })\rangle $ determines spectral properties of
the magnetization.  Namely this situation is realized in magnetic
systems subject to an external regular and random magnetic field. In
both cases we assume that the characteristic wave length of the
electromagnetic field is much larger than the linear size of the
system.

In two-level systems the random field can be realized by an
environment, for example, by thermal phonons or by magnetic field of
nuclei. In such a situation different two-level systems feel different
random fields. Therefore, in the ensemble consisting of a large number
$N$ of the two-level systems the fluctuation will be suppressed
proportional to $1/\sqrt{N}$. However, the two-time correlation
function describes the linear response of such a system to a small
perturbation $\delta {\bf h}(t)$ (identical for all members) according
to a standard linear-response equation:
\begin{equation}
  \delta\langle s_i\rangle = \int_{-\infty}^t dt^{\prime} \ 
  \langle s_i(t)s_j(t^{\prime})\rangle \delta h_j(t^{\prime}) .
\label{response}
\end{equation}
Such a perturbation can be realized as a pulse of electric magnetic or
acoustic field. In this work we determine all these correlations under
the same assumption of fast noise.

The outline of the paper is as follows.  In Sec.~\ref{sec.fun} we
formulate the model for LZ transitions in the presence of noise we
focus on.  In Sec.~\ref{sec.noiseonly} we calculate in a general form
the time evolution of averages and autocorrelations of the Bloch
vector in the absence of the regular transition matrix element
$\Delta$.  These general findings are evaluated to obtain the averages
and fluctuations of transition probabilities first for $\Delta=0$ in
Sec.~\ref{sec.trans0} and then also for $\Delta>0$ in
Sec.~\ref{sec.trans1}.  Implications for the spectral width and
intensity of fluctuations for transitions in a gas of colliding atoms
or molecules are discussed in Sec.~\ref{sec.spect}.  We close with
concluding remarks in Sec.~\ref{sec.conc}.  Some calculational details
are collected in an appendix.

\section{Fundamental}
\label{sec.fun}

We consider the dynamics of a quantum spin $\hbS$ in the presence
of a time dependent effective magnetic field $\bB(t)$.  Absorbing
the Land\'e factor and the Bohr magneton into this field, the
Hamiltonian reads
\begin{eqnarray}
  \hat H(t) = -\bB(t) \cdot \hat \bS \ .
\label{def.H}
\end{eqnarray}
The effective field $\bB(t)= \bb(t) + \bfeta(t)$ is composed of a
regular part $\bb$ and a noisy part $\bfeta$ with zero average,
$\langle \bfeta(t)\rangle={\bf 0}$.  In the LZ theory, an avoided
level crossing is described by the regular field which may be chosen as
\begin{eqnarray}
  \bb(t) = b_z(t) \be_z + b_x \be_x
\end{eqnarray}
with a $z$ component approximately linear in time,
\begin{eqnarray}
  b_z(t) \approx \nu t, &&
  \nu \equiv \dot b_z(0),
\label{def.nu}
\end{eqnarray}
and an approximately time independent perpendicular component
$b_x=-(2/\hbar) \Delta$.  The noise is considered as Gaussian
distributed and uncorrelated in the longitudinal and transverse
directions with variances ($i,j=x,y,z$)
\begin{eqnarray}
  \langle \eta_i(t) \eta_j(t') \rangle &=& \delta_{ij} \ f_i(t-t') \ .
\end{eqnarray}
The correlation functions $f_i$ naturally are even functions.  A
further planar symmetry $f_x=f_y \equiv f_\perp$ is assumed.

Under the action of the Hamiltonian (\ref{def.H}), the dynamics of
the density matrix is equivalent to the classical equation of
motion (Bloch equation)
\begin{eqnarray}
  \dot {\mathbf{g}}(t) = - \mathbf{B}(t) \times \mathbf{g}(t)  \ .
\end{eqnarray}
For the analysis of the Bloch equation it is convenient to introduce
\begin{eqnarray}
  g_{\pm } \equiv \frac{1}{\sqrt{2}}(g_{x}\pm ig_{y})
\label{def.gpm}
\end{eqnarray}
as well as $\eta _{\pm }$, $b_{\pm }$, and $B_\pm$ by analogous
definitions.  In these terms the Bloch equation can be rewritten as
\begin{subequations}
\begin{eqnarray}
  \dot g_\pm &=& \mp i B_z g_\pm \pm i B_\pm g_z \ ,
  \\
  \dot g_z &=&  i [ B_- g_+ - B_+g_- ] \ .
\end{eqnarray}
\label{dot.g}
\end{subequations}
From these equations the longitudinal field component $B_z(t)$ can
be eliminated by going over to the interaction representation with
respect to the Hamiltonian
\begin{eqnarray}
  \hat{H}_{0}(t)=-B_{z}(t)\hat{S}_{z} \ .
\end{eqnarray}
Using the time evolution operator
\begin{eqnarray}
  \hat{U}_{0}(t)=\exp \left\{ \frac{1}{i\hbar }
    \int_{t_0}^{t}dt^{\prime }\hat{H}_{0}(t^{\prime })\right\}
\end{eqnarray}
we rewrite the density matrix (\ref{rho}) as
\begin{eqnarray}
  \hat{\rho}(t)=\frac{1}{2}[\hat{1}+\hat{U}_{0}(t)\tilde{\mathbf{g}}(t)\cdot
  \hbsig \hat{U}_{0}^{\dagger }(t)]
\end{eqnarray}
in terms of the transformed Bloch vector
\begin{eqnarray}
  \tilde{g}_{\pm }(t)&=&
  g_{\pm }(t)\exp \left\{ \pm i\int_{t_0}^{t}dt^{\prime }
    B_{z}(t^{\prime }) \right\} \ .
\label{def.gp}
\end{eqnarray}
The transformation does not affect the longitudinal component,
$\tilde{g}_{z}(t)=g_{z}(t)$. Using analogous transformations for
$\tilde \bfeta$, $\tilde\bb$ and $\tilde \bB$, the Bloch equation
simplifies to
\begin{subequations}
\begin{eqnarray}
  \dot \gp_\pm &=& \pm i \bp_\pm \gp_z \ ,
  \\
  \dot \gp_z &=& i [ \bp_- \gp_+  - \bp_+ \gp_- ] \ .
\end{eqnarray}
\label{dot.gp}
\end{subequations}

In the limit of short noise correlation time $\tau_{\mathrm{n}}$, the
correlations for the transformed noise read
\begin{subequations}
\label{eta.corr}
\begin{eqnarray}
  \langle \etap_+(t) \etap_+(t') \rangle
  = \langle \etap_-(t) \etap_-(t') \rangle&=& 0 \ ,
  \\
  \langle \etap_+(t) \etap_-(t') \rangle &=& \fp_\perp(t,t') \ ,
\end{eqnarray}
with the correlator
\begin{eqnarray}
  \fp_\perp(t,t') &\equiv& f_\perp(t-t')
  \exp \left\{ i\int_{t'}^{t}dt''
    B_{z}(t'') \right\}
  \nonumber \\
  &\approx& f_\perp(t-t') \ e^{ i \nu (t^2 - t'^2)/2} \ .
\end{eqnarray}
\end{subequations}
Since $ f_\perp(t-t')$ decays for $|t-t'|\gg \tn$ one may neglect the
contribution of $\eta _{z}(t'')$ to the time integral provided
\begin{eqnarray}
  \tau_{\mathrm{n}}^{2} \langle \eta _{z}^{2}\rangle \ll 1 \ .
\end{eqnarray}
This condition is assumed in the following. It can be considered as a
further limitation for the noise correlation time or as a limitation
for the noise amplitude. However, it is not crucial and is introduced
only for simplification. If the longitudinal noise is independent of
the transverse one, its contribution to the averaged exponent in
equation (\ref{eta.corr}) can be reduced to the Debye-Waller factor
$W(t,t^{\prime})=\exp{[-\frac{1}{2}
  \int_{t^{\prime}}^tdt_1\int_{t^{\prime}}^tdt_2 \langle\eta (t_1)\eta
  (t_2)\rangle]}$. Theory becomes less transparent and more
cumbersome, but in essence remains the same except for the range of
very strong longitudinal noise.

Although the longitudinal noise is not effective in $\fp_\perp$,
it cannot be neglected in general. From Eq.~(\ref{def.gp}) one
recognizes that $\eta _{z}$ leads to a diffusive precession of the
transverse components of $\mathbf{g}$. Therefore, the expectation
values $\langle g_{\pm }\rangle $ decay exponentially on the
dephasing time
\begin{eqnarray}
  \tau _{\phi }=\frac{2}{F_{z0}}
\label{def.tauphi}
\end{eqnarray}
with $F_{z0} \equiv \int_{-\infty }^{\infty }dt \ \langle \eta
_{z}(t)\eta _{z}(0)\rangle $. For short noise correlations, $\tau
_{\phi }\sim 1/[\tau_{ \mathrm{n}}\langle \eta _{z}^{2}\rangle ]$.
Thus, longitudinal dephasing can be neglected during the
accumulation period $|t|\lesssim \tau_{\mathrm{acc} }=1/\nu\tau_{
\mathrm{n}}$ of transverse noise for $\langle \eta _{z}^{2}\rangle
\ll \nu $. On the other hand, on sufficiently large time scales
$|t| \gg \tau_\phi$ dephasing always sets in. If the dynamics of
the system is followed over that long times, the averages of $g_x$
and $g_y$ vanish because of the random precession around the $z$
axis. Nevertheless, in this case the transverse amplitude
\begin{eqnarray}
  g_\perp \equiv (g_x^2 + g_y^2)^{1/2}
  = (2 g_+ g_-)^{1/2} = (2 \gp_+ \gp_-)^{1/2} \ ,
\end{eqnarray}
which is identical to the absolute value of the off-diagonal element
of the density matrix, can have a finite expectation value.

\section{Noisy dynamics}
\label{sec.noiseonly}

In this section, we temporarily neglect regular transitions (we set
$b_{x}=0$) and focus on transitions solely due to noise.  To solve the
equations of motion for the propagation from an initial time ${t_{0}}$
to $t>{t_{0}}$, we integrate the equations of motion (\ref{dot.gp}) to
\begin{subequations}
\begin{eqnarray}
  \gp_\pm(t) &=&  \gp_\pm(\ti) \pm i \int_\ti^t dt' \etap_\pm(t') \gp_z(t')
  \label{gp.pm}
  \\
  \gp_z(t) &=& \gp_z(\ti)  - \int_\ti^t dt_1 \int_\ti^{t_1} dt_2
    w(t_1,t_2) \gp_z(t_2)
  \nonumber \\
  &&
  - \int_\ti^t dt'  [i\etap_+(t')  \gp_-(\ti) + \cc]
\label{gp.z}
\end{eqnarray}
\end{subequations}
with the vertex pair function
\begin{eqnarray}
  w(t_{1},t_{2}) \equiv \tilde{\eta}_{+}(t_{1}) \tilde{\eta}_{-}(t_{2})
  + \cc \ .
\label{def.w}
\end{eqnarray}
Wherever ``$\cc$'' occurs, it stands for the complex conjugation of
the preceding term.

\begin{figure}[htbp]
\centering
\epsfig{file=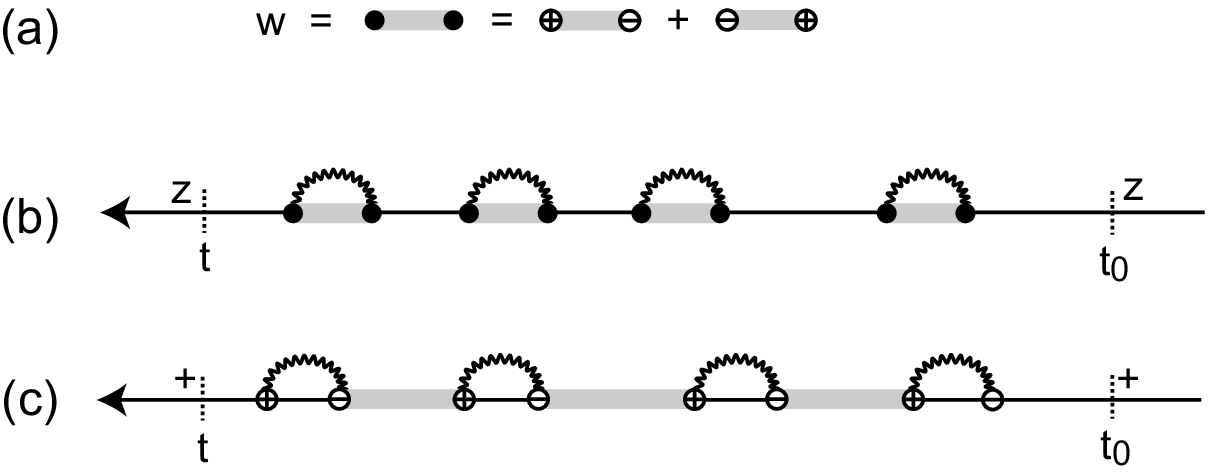,width=\figwidth}
\caption{%
  Diagrammatic representation of $w$ and contributions to $%
  G_{\alpha,\alpha}(t,{t_0})$. The time axis is chosen to point to the
  left.  An encircled $+$ or $-$ represents a noise vertex $\tilde
  \eta_\pm$. (a) A pair of vertices connected by a gray bond
  represents $-w(t_1,t_2)$. A pair of bold circles contains the sum
  over both possible polarities. (b) To leading order in small
  $\tau_{\mathrm{n}}$, only diagrams with noncrossing vertex
  contractions contribute. $G_{z,z}(t,{t_0})$ consists only of noise
  contraction (wiggly lines) within vertex pairs. (c)
  $G_{+,+}(t,{t_0})$ consists on a negative excess vertex as first
  vertex (closest to ${t_0}$) and a positive excess vertex as last
  vertex. Contractions can be performed only from pair to pair.}
\label{fig_av}
\end{figure}

Equation (\ref{gp.z}) may be fed back iteratively into itself and into
Eq.~(\ref{gp.pm}), to express $\tilde{\mathbf{g}}(t)$ entirely in
terms of $\tilde{\mathbf{g}}({t_0})$. Each term arising from this
iteration can be visualized by a diagram.  Defining the integral
series
\begin{widetext}
  \begin{eqnarray}
    W_\ti^t(t') &\equiv&  \delta(\ti-t')
    -\int_\ti^{t} dt_1 \int_\ti^{t_1} dt_2 w(t_1,t_2) \delta(t_2-t')
    \nonumber \\ &&
    + \int_\ti^{t} dt_1 \int_\ti^{t_1} dt_2
    \int_\ti^{t_2} dt_3 \int_\ti^{t_3} dt_4
    w(t_1,t_2) w(t_3,t_4) \delta(t_4-t')
    - \dots
  \end{eqnarray}
one obtains
\begin{subequations}
\begin{eqnarray}
  \gp_z(t) &=& \int_{-\infty}^\infty dt' W_\ti^t(t')  \left\{\gp_z(\ti)
    - \int_\ti^{t'} dt'' [i\etap_+(t'') \gp_-(\ti) + \cc] \right\} \ ,
  \label{gz.iterated}
  \\
  \gp_\pm(t) &=& \gp_\pm(\ti) \pm i \int_\ti^t dt' \etap_\pm(t')
  \int_{-\infty}^\infty dt'' W_\ti^{t'}(t'')  \left\{\gp_z(\ti)
    - \int_\ti^{t''} dt''' [i\etap_+(t''') \gp_-(\ti) + \cc] \right\} \ .
\end{eqnarray}
\label{iterated}
\end{subequations}
\end{widetext}
In the diagrammatic representation, every factor $W$ corresponds to an
arbitrary number of vertex pairs $w$. Corresponding to the definition
(\ref{def.w}), two ``polarities'' have to be considered per pair (cf.
Fig.~\ref{fig_av}a).  During the time evolution (\ref{iterated}) of
$\tilde{\mathbf{g}}$, besides such paired vertices, ``excess'' vertices
can appear as first or last vertices during the evolution period.

According to the Wick theorem for Gaussian fields, the noise averaging
corresponds to all possible contractions between noise vertices. The
correlations (\ref{eta.corr}) imply that contractions can be performed
only between vertices with opposite charge (in a pictorial language,
we associate a ``charge'' $\pm 1 $ with every vertex $\tilde
\eta_\pm$). Therefore, only neutral diagrams (where the number of
vertices $\tilde \eta_+$ equals the number of vertices $\tilde
\eta_-$) do not vanish.

\subsection{Dynamics of averages}

Before addressing correlations, we analyze which diagrams contribute
to the average $\langle \tilde{\mathbf{g}}(t)\rangle $. Because of the
neutrality constraint, this average reduces to
\begin{subequations}
\begin{eqnarray}
  \langle \gp_z(t)\rangle  &=& \int_{-\infty}^\infty dt'
  \langle W_\ti^t(t') \rangle \langle \gp_z(\ti) \rangle \ ,
  \label{diff.z}
  \\
  \langle \gp_\pm(t) \rangle&=&
  \bigg\{ 1 -
    \int_\ti^t dt' \int_{-\infty}^\infty dt'' \int_\ti^{t''} dt'''
    \nonumber \\ && \times
    \langle \etap_\pm(t') W_\ti^{t'}(t'') \etap_\mp(t''') \rangle
  \bigg\} \langle \gp_\pm(\ti) \rangle \ .
\label{diff.pm}
\end{eqnarray}
\end{subequations}
In factorizing the expectation values we assume that noise before and
after ${t_{0}}$ is statistically independent.

To leading order in $\tau _{\mathrm{n}}\ll \tau _{\mathrm{acc}}$, only
contractions between neighboring vertices (in time order) contribute.
In $\langle \tilde{g}_{z}(t)\rangle $, all contractions are performed
within the pair functions $w(t,t^{\prime })$ (cf.
Fig.~\ref{fig_av}b), which (for $t-{t_{0}}\gg \tau _{\mathrm{n} }$)
give rise to factors
\[
\int_{t_{0}}^{t}dt^{\prime }\langle w(t,t^{\prime })\rangle =F_\perp(\nu t)
\]
with the Fourier transform
\begin{equation}
  F_\perp(\Omega ) \equiv \int_{-\infty }^{\infty }d\tau \
  \cos (\Omega \tau ) f_\perp(\tau) \ .
  \label{Fourier1}
\end{equation}

Returning to Eq. (\ref{diff.z}), we find that the series can be easily
summed up to
\begin{subequations}
\begin{eqnarray}
  \langle \gp_z(t)\rangle &=&
  G_{z,z}(t,\ti) \langle \gp_z(\ti) \rangle
\end{eqnarray}
with
\begin{eqnarray}
  G_{z,z}(t,t_{0}) &\equiv& \int_{-\infty }^{\infty }dt^{\prime }\langle
  W_{t_{0}}^{t}(t^{\prime })\rangle
  \nonumber \\
  &=& \exp \left\{ -\int_{t_{0}}^{t}dt^{\prime } F_\perp
    (\nu t^{\prime })\right\}  \ .
\end{eqnarray}
\label{prop-z}
\end{subequations}
This result was obtained in the Ref.~\cite{PS1} via the
differential equation
\begin{eqnarray}
  \frac{d\langle \gp_{z}(t)\rangle }{dt}=-F_\perp(\nu t)
  \langle \gp_z(t)\rangle
\label{diff.gz}
\end{eqnarray}
which immediately follows from the integral formula (\ref{prop-z}) and
confirms the statistical independence of the events before and after
some moment of time $t$ on a time scale much larger than $\tn$.

Likewise, to leading order in $\tau _{\mathrm{n}}\ll \tau
_{\mathrm{acc}}$, $\langle \tilde{g}_{+}(t)\rangle $ consists only of
contractions linking neighboring pairs due to the presence of excess
vertices (cf.  Fig.~\ref{fig_av}c).  Such a contraction leads to a
factor
\[
\int_{t_{0}}^{t}dt^{\prime }\langle \tilde{\eta}_{\pm }(t)\tilde{\eta}_{\mp
}(t^{\prime })\rangle =F_\perp^{\pm }(\nu t)
\]
with
\[
F_\perp^{\pm }(\Omega ) \equiv
\int_{0}^{\infty }d\tau\ e^{\pm i\Omega \tau }f_\perp(\tau )  \ .
\]
Apparently, $F_\perp=F_\perp^{+}+F_\perp^{-}$ and
$F_\perp^{+}={F_\perp^{-}}^{*}$ because of the even parity of the
noise correlator $f_\perp(t)$.  Hence, resummation of diagrams gives
\begin{eqnarray}
    \langle \gp_+(t)\rangle &=&
  G_{+,+}(t,\ti) \langle \gp_+(\ti) \rangle
\end{eqnarray}
with
\begin{eqnarray}
  G_{+,+}(t,t_{0}) &\equiv& 1-\int_{t_{0}}^{t}dt^{\prime }\int_{-\infty
  }^{\infty }dt^{\prime \prime }\int_{t_{0}}^{t^{\prime \prime }}dt^{\prime
    \prime \prime }
  \nonumber \\ && \times
\langle \tilde{\eta}_{\pm }(t^{\prime })W_{t_{0}}^{t^{\prime
}}(t^{\prime \prime })\tilde{\eta}_{\mp }(t^{\prime \prime \prime })\rangle
 \nonumber \\
&=&\exp \left\{ -\int_{t_{0}}^{t}dt^{\prime }F_\perp^{\pm }(\nu t^{\prime
})\right\} \ .
\label{perp}
\end{eqnarray}

In summary, the dynamics of averages can brought into the compact form
($\alpha =z,\pm $)
\begin{subequations}
\begin{eqnarray}
  \langle \gp_\alpha(t)\rangle &=&
  G_{\alpha,\alpha}(t,\ti) \langle \gp_\alpha(\ti) \rangle \ ,
  \\
  G_{\alpha,\alpha}(t,\ti) & \equiv&
  e^{-[\vartheta_\alpha(t)-\vartheta_\alpha(\ti)]}
\end{eqnarray}
\label{gen}
\end{subequations}
by means of noise integrals
\begin{subequations}
\begin{eqnarray}
  \vartheta _{\pm}(t) &\equiv& \int_{-\infty }^{t}dt^{\prime }
  F_\perp^{\pm}(\nu t^{\prime }) \ ,
  \\
  \vartheta _{z}(t) &\equiv& \int_{-\infty }^{t}dt^{\prime }
  F_{\perp}(\nu t^{\prime })  \ .
\end{eqnarray}
\end{subequations}
Note that the subscripts of $\vartheta$ relate to the components of
the Bloch vector rather than to the noise components.

\subsection{Dynamics of pair correlations}

Using the diagrammatic approach established above, we now proceed to
calculate pair correlations of the Bloch vector for $b_x=0$.

\subsubsection{Reduction to equal-time correlators}

We start by realizing that
\begin{equation}
  \langle \tilde{g}_{\alpha }(t)\tilde{g}_{\beta }(\bar{t})\rangle
  =e^{-[\vartheta _{\alpha }(t)-\vartheta _{\alpha }(\bar{t})]}
  \langle \tilde{g}_{\alpha }(\bar{t})\tilde{g}_{\beta }(\bar{t})\rangle
  \label{two-time}
\end{equation}
since noise is uncorrelated before and after $\bar{t}$, assuming (without
loss of generality) that $t>\bar{t}$. Thus we need to calculate only
equal-time correlators
\[
\tilde{C}_{\alpha \beta }(t) \equiv
\langle \tilde{g}_{\alpha }(t)\tilde{g}_{\beta}(t)\rangle  \ .
\]
The correlations of principal interest are
\begin{subequations}
\begin{eqnarray}
  \langle g_z (t) g_z(t) \rangle &=& \langle \gp_z (t) \gp_z(t) \rangle \ ,
  \\
  \langle g_+ (t) g_-(t) \rangle &=& \langle \gp_+ (t) \gp_-(t) \rangle \ .
\end{eqnarray}
\end{subequations}
They are invariant under the transformation to the interaction
representation.

We describe the evolution of the equal-time correlators
\[
\tilde{C}_{\alpha \beta }(t)=\sum_{\gamma \delta }G_{\alpha \beta ,\gamma
\delta }(t,{t_{0}})\tilde{C}_{\gamma \delta }({t_{0}})
\]
in terms of propagators $G_{\alpha \beta ,\gamma \delta }(t,{t_{0}})$.
In the diagrammatic representation, these propagators consist only of
ladder-like diagrams: The times axes $t$ and $\bar{t}$ represent the
legs of the ladder. Noise contractions between the sides are the
rungs.  A contribution of any diagram with intersecting or overlapping
lines is smaller than that of the main sequence by a factor of the
order $\tn/\tau_{\mathrm{acc}}$.  Since only ``neutral'' diagrams
survive averaging, many propagators vanish. The surviving propagators
satisfy ``charge conservation''
\[
\chi (\alpha )+\chi (\beta )=\chi (\gamma )+\chi (\delta )
\]
with the charge function
\[
\chi (\alpha ) \equiv \left\{
\begin{array}{ccl}
1 & \text{for} & \alpha =+ \ ,\\
0 & \text{for} & \alpha =z \ ,\\
-1 & \text{for} & \alpha =- \ ,
\end{array}
\right.
\]
which now refers to propagator indices, not to noise indices. At the
initial time ${t_{0}}$, the propagator charge is opposite to the
charge of the first excess vertex. At the final time, the propagator
charge coincides with the charge of the last excess vertex.

Among the nonvanishing propagators, several propagators are mutually
dependent. In particular, there is a trivial symmetry $G_{\alpha \beta
  ,\gamma \delta }=G_{\beta \alpha ,\delta \gamma }$ corresponding to
the exchange of the legs of the ladder.  Additional relations follows
from the time-reversal invariance or the complex conjugation, such as
$G_{z-,z-}=G_{z+,z+}^{*}$.

We therefore find the time evolution of the equal-time correlators
captured by the relations
\begin{subequations}
\label{C.prop}
\begin{eqnarray}
  \Cp_{zz}(t) &=&
  G_{zz,zz}(t,\ti) \Cp_{zz}(\ti)
  \nonumber \\ && +
  2 G_{zz,+-}(t,\ti) \Cp_{+-}(\ti) \ ,
  \\
  \Cp_{+-}(t) &=&
  G_{+-,zz}(t,\ti) \Cp_{zz}(\ti)
  \nonumber \\ && +
  [G_{+-,+-}(t,\ti)
  \nonumber \\ &&
  + G_{+-,-+}(t,\ti) ]\Cp_{+-}(\ti) \ ,
  \\
  \Cp_{++}(t)  &=&
  G_{++,++}(t,\ti) \Cp_{++}(\ti) \ ,
  \\
  \Cp_{z+}(t) &=&
  [G_{z+,z+}(t,\ti)
  \nonumber \\ &&
  + G_{z+,+z}(t,\ti)] \Cp_{z+}(\ti) \ .
\end{eqnarray}
\end{subequations}
It is important to realize that the dynamics of the ``neutral''
correlations $\tilde{C}_{z,z}$ and $\tilde{C}_{+,-}$ is decoupled from
the ``non-neutral'' $\tilde{C}_{z,\pm }$ and $\tilde{C}_{\pm ,\pm }$.
This implies that the latter correlations will not be generated by
noise if they are absent initially.  However, such correlations can be
generated in general by the transverse components of the magnetic
field.

For the convenience of the reader we anticipate the results of our
subsequent explicit calculation of the propagators:
\begin{subequations}
\begin{eqnarray}
  G_{zz,zz}(t,\ti) &=& \frac 13 (1+2 e^{-3 \theta_z}) \ ,
   \label{G.zz.zz}
   \\
   G_{+-,zz}(t,\ti) &=& \frac 13  (1-e^{-3 \theta_z}) \ ,
  \label{G.+-.zz}
  \\
  G_{zz,+-}(t,\ti) &=&  G_{+-,zz}(t,\ti) \ ,
  \label{G.zz.+-}
  \\
  G_{+-,+-}(t,\ti) &=& \frac16 (2 + 3 e^{-\theta_z} + e^{-3\theta_z} ) \ ,
  \label{G.+-.+-}
  \\
  G_{+-,-+}(t,\ti) &=& \frac16 (2-3 e^{-\theta_z} + e^{-3\theta_z} ) \ ,
  \label{G.+-.-+}
\end{eqnarray}
\end{subequations}
introducing the abbreviation
\begin{eqnarray}
\theta_\alpha \equiv \vartheta_\alpha(t)-\vartheta_\alpha({t_0}) \ .
\end{eqnarray}

These explicit calculations are based on integral equations which we
now derive and solve for the propagators $G_{zz,zz}$ and $G_{zz,+-}$
as two representative examples. The diagrammatic derivation of the
other propagators is deferred to the appendix.  We refrain from
presenting the expressions for $G_{++,++}$, $G_{z+,z+}$, and
$G_{z+,+z}$ since they are not relevant for our purposes.

\subsubsection{Propagator $G_{zz,zz}$}

We start with the propagator for the longitudinal correlator
\[
G_{zz,zz}(t,{t_{0}})=\int_{-\infty }^{\infty }dt^{\prime
}\int_{-\infty }^{\infty }d\bar{t}^{\prime }\langle
W_{t_{0}}^{t}(t^{\prime })W_{t_{0}}^{t}( \bar{t}^{\prime })\rangle \ .
\]
This expression follows directly from the contribution to the
autocorrelation of Eq.~(\ref{gz.iterated}) stemming from the initial
value $\gp_z(t_0)$.

\begin{figure}[tbph]
\centering
\epsfig{file=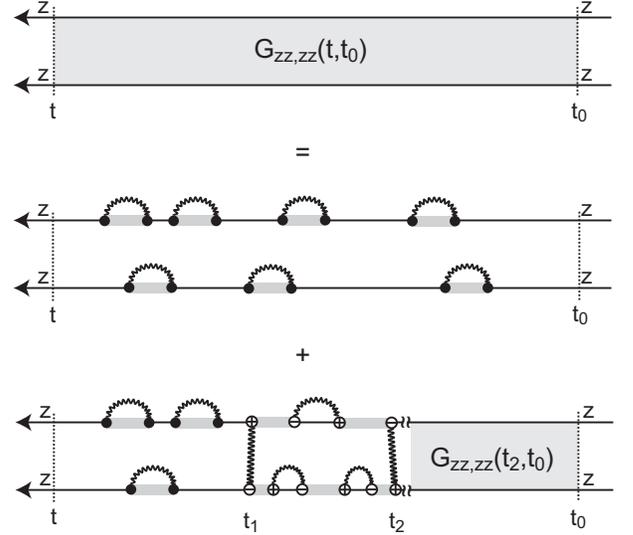,width=\figwidth}
\caption{%
  Diagrams contributing to $G_{zz,zz}(t,{t_{0}})$. First, there is the
  ``ladder'' without rungs. The diagrams with rungs can be represented
  as propagators $G_{z,z}(t,t_{1})$ on both legs of the ladder,
  following the ring contraction between the last rungs as times
  $t_{1}$ and $t_{2}$. Prior to $t_{2}$ additional leg propagators and
  rings may occur. In the Bethe-Salpeter equation, these additional
  contraction sum up to $ G_{zz,zz}(t_{2},{t_{0}})$.}
\label{figzzzz}
\end{figure}

To leading order in $\tn/\tacc$, the contributing diagrams have the
ladder structure mentioned previously (cf. Fig. \ref{figzzzz}).  On
leg segments between two neighboring rungs of the ladder, an arbitrary
number of noncrossing vertex contractions can be performed.  Since
there are no final excess vertices close to time $t$, all contractions
between the last rung (at time $t_1$) and the final time $t$ are
intra-pair contractions.  Between the last rung and the last but one
rung, all contractions are inter-pair contraction, forming a ring
structure limited by time $t_1$ and $t_2$.  Then, the next
contractions on the legs prior to $t_2$ can be intra-pair contractions
again.

Since all possible diagrams prior to $t_2$ have the same structure as
the diagrams prior to $t$, the propagator satisfies a Bethe-Salpeter
integral equation:
\begin{eqnarray}
  G_{zz,zz}(t,{t_{0}}) &=&
  G_{z,z}^{2}(t,{t_{0}})+2\int_{t_{0}}^{t}dt_{1}
  \int_{t_{0}}^{t_{1}}dt_{2}G_{z,z}^{2}(t,t_{1})
   \nonumber \\
   &&\times f(\nu t_{1})G_{+,+}(t_{1},t_{2})G_{-,-}(t_{1},t_{2})f(\nu t_{2})
   \nonumber \\
   &&\times G_{zz,zz}(t_{2},{t_{0}}) \ .
\label{Gzzzz-integral}
\end{eqnarray}
The first term represents the diagram without rungs. In the second
term, the last pair of rungs at times $t_{1}$ and $t_{2}$ is separated
from possible additional pairs prior to $t_{2}$. On the ring between
$t_1$ and $t_2$, the intra-pair noise contractions are ``oriented''.
The explicit factor $2$ arises from the two possible orientations.

To simplify the integral equation, it is convenient to substitute the
time variables by variables $y \equiv \vartheta _{z}(t)-\vartheta
_{z}({t_{0}})$ and $y_{i} \equiv \vartheta _{z}(t_{i})-\vartheta
_{z}({t_{0}})$.  This leads to
\begin{eqnarray}
  G_{zz,zz}(y,0) &=&e^{-2y}+2\int_{0}^{y}dy_{1}\int_{0}^{y_{1}}dy_{2}
  \nonumber \\
  &&\times e^{-2(y-y_{1})}e^{-(y_{1}-y_{2})}G_{zz,zz}(y_{2},0) \ .
\end{eqnarray}
This integral equation can now be transformed into a differential equation
in two steps:
\begin{subequations}
\begin{eqnarray}
  \frac d{dy} [e^{2y} G_{zz,zz}(y,0)] &=&  2 \int_0^{y} dy_2
  \ e^{y +y_2}
  \nonumber \\ && \times
  G_{zz,zz}(y_2,0) \ ,
  \label{zz.zz.step1}
  \\
  \frac d{dy} \{ e^{-y}\frac d{dy} [e^{2y} G_{zz,zz}(y,0)]\} &=&
  2 e^{y} G_{zz,zz}(y,0) \ .
\end{eqnarray}
\end{subequations}
Performing the derivatives, the last equation simplifies to
\[
\frac{d^{2}}{dy^{2}}G_{zz,zz}(y,0)+3\frac{d}{dy}G_{zz,zz}(y,0)=0  \ .
\]
This differential equation must be solved with the initial conditions
\begin{subequations}
\begin{eqnarray}
  G_{zz,zz}(0,0) &=&1 \ ,
  \\
  \frac {d}{dy} G_{zz,zz}(y,0)|_{y=0} + 2  G_{zz,zz}(0,0) &=& 0 \ .
\end{eqnarray}
\end{subequations}
The first one expresses the trivial fact that there is no time
evolution over a vanishing time interval and follows from equation
(\ref{Gzzzz-integral}). The second one stems directly from Eq.
(\ref{zz.zz.step1}). Both determine a surprisingly simple form of the
final result (\ref{G.zz.zz}) for $G_{zz,zz}(t,{t_{0}})$.

Considering an initial state with $\tilde{\mathbf{g}}({t_0})=\tilde
g_z({t_0} ) \mathbf{e}_z$, the conservation of $\mathbf{g}^2$ implies
the normalization
\begin{eqnarray}
G_{zz,zz} + 2 G_{+-,zz} = 1 \ .
\end{eqnarray}
This implies the result (\ref{G.+-.zz}) for $G_{+-,zz}(t,{t_0})$ which
is derived also diagrammatically in the appendix.

\subsubsection{Propagator $G_{zz,+-}$}
\label{sec.G.zzpm}

\begin{figure}[tbph]
\centering
\epsfig{file=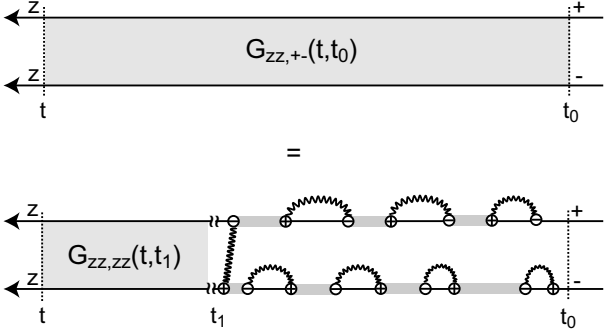,width=\figwidth}
\caption{%
  The propagator $G_{zz,+-}$ consists of diagrams in which the first
  vertices are excess vertices. They can be contracted with a certain
  number of pairs until the first rung occurs at time $t_{1}$. The
  contraction of all later pairs yields $G_{zz,zz}(t,t_{1})$.}
\label{figzzpm}
\end{figure}

As an example for the diagrammatic calculation of the other propagators, we
consider here
\begin{eqnarray}
  G_{zz,+-}(t,t_{0}) &=&\int_{-\infty }^{\infty }dt^{\prime}
  \int_{t_{0}}^{t^{\prime }}dt^{\prime \prime }
  \int_{-\infty }^{\infty }d\bar{t}^{\prime }
  \int_{t_{0}}^{\bar{t}^{\prime }}d\bar{t}^{\prime \prime }
  \nonumber \\
  &&\times \langle W_{t_{0}}^{t}(t^{\prime })
  W_{t_{0}}^{t}(\bar{t}^{\prime })
  \tilde{\eta}_{-}(t^{\prime \prime })
  \tilde{\eta}_{+}(\bar{t}^{\prime \prime})\rangle \ .
\end{eqnarray}
This expression follows directly from the contribution to the
autocorrelation of Eq.~(\ref{gz.iterated}) stemming from the initial
value $\gp_\pm(t_0)$.  Its diagrams (cf. Fig.~\ref{figzzpm}) have no
excess vertices close to the final time $t$ but opposite excess
vertices close to the initial time $t_0$.  Therefore, close to $t_0$
only inter-pair contractions are possible which are connected through
the first rung at the time $t_1$.  All possible contractions between
$t_1$ and $t$ have the same structure as the diagrams of $G_{zz,zz}$.
Therefore, the diagrams can be summed implicitly to
\begin{eqnarray}
G_{zz,+-}(t,t_{0}) &=&\int_{t_{0}}^{t} dt_1 G_{zz,zz}(t,t_{1})f(\nu t_{1})
\nonumber \\
&&\times G_{+,+}(t_{1},{t_{0}})G_{-,-}(t_{1},{t_{0}}) \ .
\end{eqnarray}
After a transition to variables $y$ the integral can be performed
explicitly and yields $G_{zz,+-}=G_{+-,zz}$.

The normalization condition
\[
G_{zz,+-}(t,{t_{0}})+G_{+-,+-}(t,{t_{0}})+G_{+-,-+}(t,{t_{0}})=1  \ ,
\]
which expresses the conservation of $\mathbf{g}^{2}$ for an initial
transverse state, determines already the sum of the propagators
$G_{+-,+-}(t,{t_{0}})$ and $G_{+-,-+}(t,{t_{0}})$ which enter
Eqs.~(\ref{C.prop}) only in this combination.  Nevertheless, both
propagators $G_{+-,+-}(t,{t_{0}})$ and $G_{+-,-+}(t,{t_{0}})$ are
calculated separately in the appendix.

\section{Transitions produced by noise only}
\label{sec.trans0}

On the basis of the propagators derived in the previous section we now
evaluate transition probabilities and their fluctuations, still
focusing on the subcase $b_x=0$.  We consider an arbitrary initial
state at $t_{0}=-\infty $ with arbitrary Bloch vector
$\mathbf{g}(-\infty )$. We are interested in the final time
$t=\infty$, where the average is given by \cite{PS2}
\begin{subequations}
\begin{eqnarray}
  \langle g_z(\infty) \rangle &=& e^{-\Theta} g_z(-\infty) \ ,
  \\
  \langle \gp_\pm(\infty) \rangle &=& e^{-\Theta/2} \ \gp_\pm(-\infty)  \ .
\end{eqnarray}
\end{subequations}
according to Eqs.~(\ref{gen}).  Since the noise correlators are even
functions of time differences, only the real parts of $\vartheta
_{\alpha }$ contribute to the difference $\vartheta _{\alpha }(\infty)
- \vartheta _{\alpha }(-\infty)$. We have introduced
\[
\Theta_\alpha \equiv \vartheta _{z}(\infty )=
\frac{\pi}{\nu}  \langle \eta _{x}^{2}+\eta_{y}^{2}\rangle
\]
for abbreviation. It is important to realize that the decay of the
averaged Bloch vector depends only on the \textit{instantaneous} noise
correlation.

Although the final slow amplitudes $\tilde{g}_{\pm}(\infty)$ will
be finite if the initial amplitudes $\tilde{g}_{\pm}(-\infty)$
were finite, the original amplitudes vanish,
\begin{eqnarray}
   \langle g_\pm(\infty) \rangle =0
\end{eqnarray}
due to the diffusive random precession leading to a decay on the
dephasing time $\tau_\phi$ given in Eq. (\ref{def.tauphi}).  However,
the transverse amplitude $g_\perp$ decays like $\gp_\pm$,
\begin{eqnarray}
  \langle g_\perp(\infty) \rangle &=& e^{-\Theta/2} \ g_\perp(-\infty)  \ .
\end{eqnarray}

For the variances we obtain -- suppressing the obvious time arguments
$(\infty ,-\infty )$ of the propagators --
\begin{subequations}
\begin{eqnarray}
  \langle g_z^2(\infty) \rangle   &=&
  G_{zz,zz}  g_z^2(-\infty)
  + G_{zz,+-}  g_\perp^2(-\infty)
  \nonumber \\ &=&
  \frac 13 \{ \bg^2  +  e^{-3\Theta}
  [2 g_z^2(-\infty) - g_\perp^2(-\infty) ]\} \ ,
  \\
  \langle g_\perp^2(\infty) \rangle &=&
  2 G_{+-,zz} g_z^2(-\infty)
  \nonumber \\ &&
  + [G_{+-,+-}  + G_{+-,-+} ]g_\perp^2(-\infty)
  \nonumber \\ &=&
  \frac 13 \{2 \bg^2 - e^{-3\Theta}
  [2 g_z^2(-\infty) - g_\perp^2(-\infty) ]\} \ ,
\end{eqnarray}
\end{subequations}
and for the fluctuations
\begin{subequations}
\begin{eqnarray}
  \langle [\delta g_z(\infty)]^2 \rangle &\equiv&
  \langle g_z^2(\infty) \rangle -   \langle g_z(\infty) \rangle^2
  \nonumber \\
  &=& \frac 13 \{ \bg^2 +  [2 e^{-3\Theta} - 3 e^{-2\Theta}]
  g_z^2(-\infty)
  \nonumber \\ &&
  - e^{-3\Theta} g_\perp^2(-\infty)  \} \ ,
  \\
  \langle [\delta g_\perp(\infty)]^2 \rangle &\equiv&
  \langle g_\perp^2(\infty) \rangle - \langle g_\perp(\infty) \rangle^2
  \nonumber \\
  &=&
  \frac 13 \{2\bg^2 - 2 e^{-3\Theta} g_z^2(-\infty)
  \nonumber \\ &&
  +[e^{-3\Theta} - 3 e^{-\Theta}] g_\perp^2(-\infty)\} \ .
\end{eqnarray}
\end{subequations}

For systems prepared initially in a state with a well defined $
\mathbf{g}$, the correlations $C_{z\pm }$ and $C_{\pm \pm }$ do not
necessarily vanish.  They are not explicitly considered here since
they do not couple to $C_{zz}$ and $ C_{+-}$ according to
Eqs.~\ref{C.prop}.  Furthermore, an long time scales, phase diffusion
due to $\eta_z$ annihilates these non-neutral correlations anyway.

To discuss the statistics of transitions between the two levels, we
consider the system to be located initially in the state
$|{\uparrow}\rangle $, i.e. $\mathbf{g}(-\infty )=\mathbf{e}_{z}$.  A
measurement which detects at time $t$ whether the system is in the
state $|{\uparrow}\rangle $ or $|{\downarrow}\rangle $ yields the
probabilities
\begin{eqnarray}
  \Pi_{\uparrow \uparrow/\uparrow\downarrow}(t)  &=&
  \textrm{ Tr } \frac 12 (\hat 1 \pm \hat\sigma_z) \hat \rho(t)
  = \frac{1}{2}[1\pm g_{z}(t)] \ .
\end{eqnarray}
The trace accounts for the average over quantum fluctuations for a
\textit{given} realization of noise.  The additional averaging over
the noise leads to the final expectation values
\begin{subequations}
\label{prob.nonoise}
\begin{eqnarray}
  \langle \Pi _{\uparrow \uparrow }(\infty )\rangle
  =\frac{1}{2}[1+e^{-\Theta } ],~
  \langle\Pi _{\uparrow \downarrow }(\infty)\rangle
  =\frac{1}{2}[1-e^{-\Theta }] \ .
\end{eqnarray}
Since the probabilities $\Pi_{\uparrow \uparrow/\uparrow\downarrow}$
depend on the noise, they are fluctuating quantities themselves.
Their fluctuations are
\begin{eqnarray}
\langle [\delta \Pi _{\uparrow \downarrow }(\infty )]^{2}\rangle =
\frac{1}{12}(1-3e^{-2\Theta }+2e^{-3\Theta }) \ .
\end{eqnarray}
\end{subequations}
Remarkably, the fluctuations of probabilities in general have the same
order of magnitude as average values. They are weak only if the noise
is weak, i.e., $\langle [\delta \Pi _{\uparrow \downarrow }(\infty
)]^{2}\rangle \approx \Theta ^{2}/4$ for $\Theta \ll 1$.  In the
opposite limiting case $\Theta \gg 1$ of strong noise the two levels
become equipopulated, $\langle \Pi _{\uparrow \uparrow }(\infty
)\rangle =\langle\Pi _{\uparrow \downarrow }(\infty
)\rangle=\frac{1}{2}$, and the fluctuation is equal to $\langle
[\delta \Pi _{\uparrow \downarrow }(\infty )]^{2}\rangle
=\frac{1}{12}$.

\section{LZ system with noise}
\label{sec.trans1}

In this section we assume that not only the noise, but also the
regular part of the Hamiltonian has a finite nondiagonal matrix
element $\Delta $ corresponding to a finite transverse component of
the regular magnetic field.  As it was shown in Ref.~\cite{PS1},
if the noise is fast, there exists a well-defined time separation for
the transition due to the noise and due to the regular part of the
Hamiltonian. Consider matching at times $\pm \tx$ with
$t_{\mathrm{LZ}}\ll \tx \ll \tau _{\mathrm{acc}}$. The time evolution
may be decomposed into three intervals. (i) From $t=-\infty $ to
$t=-\tx$ one may neglect $ \Delta $. The transition is only under the
influence of noise.  Since $\tx \ll \tau _{\mathrm{acc}}$, this time
evolution is approximately the same as from $t=-\infty $ to $t=0$ in
the absence of $\Delta $. (ii) From $ t=-\tx$ to $t=\tx$ the LZ
transition occurs. Since $\tx \gg \tau _{\mathrm{LZ}}$, time evolution
is approximately the same as from $t=-\infty $ to $t=\infty $ in the
absence of noise. (iii) From $t=\tx$ to $t=\infty $ the situation is
again analogous to the regime (i).

We represent the time evolution operator as
\begin{eqnarray}
  \hat U(t,{t_0}) = \hat U_0(t) \hat {\tilde U}(t,{t_0})
  \hat U_0^\dagger({t_0})
\end{eqnarray}
and approximate
\begin{subequations}
\begin{eqnarray}
  \hUp(\infty,-\infty) &=& \hUp_{\rm iii} \hUp_{\rm ii} \hUp_{\rm i} \ ,
  \\
  \hUp_{\rm i} &=& \hUp(0,-\infty)  \textrm{ for } \Delta=0 \ ,
  \\
  \hUp_{\rm ii} &=&  \hUp(\infty,-\infty)   \textrm{ for } \bfeta=0 \ ,
  \\
  \hUp_{\rm iii} &=& \hUp(\infty,0)  \textrm{ for } \Delta=0
\end{eqnarray}
\end{subequations}
for the time evolution operator acting on the slow vector.

We assume that at $t=-\infty $ the density matrix is diagonal (complete
decoherence), i.e.,
\[
\mathbf{g}(-\infty )=g_{z}(-\infty )\mathbf{e}_{z} \ .
\]
Then
\[
\langle g_{z}(-\infty )g_{z}(-\infty )\rangle =g_{z}^{2}(-\infty )
\]
is the only non-vanishing correlator at $t=-\infty $.  Since $\tn \ll
\tx$, the noise in intervals (i) and (iii) is statistically
independent and the averaging can be performed separately for both
intervals.

\subsection{Interval (i)}

Neglecting $\Delta$ in the first time interval, we find at $t=-\tx$
the averages
\begin{subequations}
\begin{eqnarray}
  \langle \gp_z(-\tx) \rangle &=& e^{-\Theta/2} g_z(-\infty) \ ,
  \\
  \langle \gp_\pm(-\tx) \rangle &=& 0 \ ,
\end{eqnarray}
\end{subequations}
and correlations
\begin{subequations}
\begin{eqnarray}
  \Cp_{zz}(-\tx)  &=&
  G_{zz,zz}(0,-\infty)  g_z^2(-\infty)
  \nonumber \\
  &=&  \frac 13 ( 1 + 2 e^{-3\Theta/2}) g_z^2(-\infty) \ ,
  \\
  \Cp_{+-}(-\tx) &=&
  G_{+-,zz}(0,-\infty)  g_z^2(-\infty)
  \nonumber \\
  &=& \frac 13 ( 1-e^{-3\Theta/2})  g_z^2(-\infty) \ .
\end{eqnarray}
\end{subequations}
According to Eqs.~(\ref{C.prop}) all other independent pair
correlations vanish at $t=-\tx$.

\subsection{Interval (ii)}

Time evolution at intermediate times from $-\tx$ to $\tx$ is given by
\[
\hat{U}_{\mathrm{ii}}=\hat{U}_{\mathrm{LZ}} \ .
\]
Hence the density matrix evolves according to
\[
\hat{\rho}(\tx)=\hat{U}_\LZ \hat{\rho}(-\tx)
\hat{U}_\LZ^{\dagger}
\]
which implies the transformation
\[
\tilde{\mathbf{g}}(\tx)\cdot \hbsig=
\hat{U}_{\mathrm{LZ}}\tilde{\mathbf{g}}(-\tx) \cdot
\hbsig {\hat{U}_{\mathrm{LZ}}^{\dagger }}
\]
for the transformed Bloch vector $\bgp$.  The explicit transformation
of its components reads
\begin{subequations}
\begin{eqnarray}
  \left(
    \begin{array}{c}
      \gp_+(\tx) \\ \gp_z(\tx) \\ \gp_-(\tx)
    \end{array}
  \right)
  &=&
  \bU^\LZ \cdot
  \left(
    \begin{array}{c}
      \gp_+(-\tx) \\ \gp_z(-\tx) \\ \gp_-(-\tx)
    \end{array}
  \right)
\end{eqnarray}
with the rotation matrix
\begin{eqnarray}
  \bU^\LZ &=&
  \left(
  \begin{array}{ccc}
    {a^*}^2 & -\sqrt2 a^* b^* & - {b^*}^2
    \\
    \sqrt2 a^* b &|a|^2-|b|^2 & \sqrt2 a b^*
    \\
    -b^2 & -\sqrt2 ab & a^2
  \end{array}
  \right) \ .
\end{eqnarray}
\end{subequations}
From these relations we obtain
\begin{subequations}
\begin{eqnarray}
  \langle \gp_z(\tx) \rangle &=& U^\LZ_{zz} \ \langle \gp_z(-\tx) \rangle
  \nonumber \\
  &=& (|a|^2-|b|^2) e^{-\Theta/2} g_z(-\infty) \ ,
  \\
  \langle \gp_+(\tx) \rangle &=& U^\LZ_{+z} \ \langle \gp_z(-\tx) \rangle
  \nonumber \\
  &=& - \sqrt2 a^* b^* e^{-\Theta/2} g_z(-\infty)
\end{eqnarray}
\end{subequations}
for the averages.  Analogously, correlations are transformed according
to
\begin{eqnarray}
\label{LZ.trans.C}
\tilde{C}_{\alpha \beta }(\tx)=\sum_{\gamma \delta }U^\LZ_{\alpha
\gamma }U^\LZ_{\beta \delta }\tilde{C}_{\gamma \delta }(-\tx) \ .
\end{eqnarray}
Since we are ultimately interested only in the correlations
$\Cp_{zz}(\infty)$ and $\Cp_{+-}(\infty)$ and since the time evolution
in interval (iii) again follows Eqs.~(\ref{C.prop}), we need to
evaluate at time $t=\tx$ only the two correlations
\begin{subequations}
\begin{eqnarray}
  \Cp_{zz}(\tx) &=&
  (U^\LZ_{zz})^2 \Cp_{zz}(-\tx)
  \nonumber \\ &&
  + 2 U^\LZ_{z+} U^\LZ_{z-} \Cp_{+-}(-\tx)
  \nonumber \\
  &=& \frac 13 [1 +2(1- 6 |a|^2|b|^2) e^{-3\Theta/2}]  g_z^2(-\infty) \ ,
  \\
  \Cp_{+-}(\tx) &=&
  U^\LZ_{+z} U^\LZ_{-z} \Cp_{zz}(-\tx)
  \nonumber \\ &&
  + [U^\LZ_{++} U^\LZ_{--}
  + U^\LZ_{+-} U^\LZ_{-+}] \Cp_{+-}(-\tx)
  \nonumber \\
  &=& \frac13 [1-(1-6 |a|^2 |b|^2) e^{-3\Theta/2}]  g_z^2(-\infty) \ .
\end{eqnarray}
\end{subequations}
As a check of these results, we verify the normalization relation
$\tilde{C}_{zz}(\tx)+2 \tilde{C}_{+-}(\tx)=g_{z}^{2}(-\infty)$.

\subsection{Interval (iii)}

As already mentioned before, the time evolution in interval (iii) is
absolutely analogous to the evolution in interval (i). We therefore
obtain
\begin{subequations}
\begin{eqnarray}
  \langle \gp_z(\infty) \rangle &=&
  G_{z,z}(\infty,0) \langle \gp_z(\tx) \rangle
  \nonumber \\
  &=& (|a|^2-|b|^2) e^{-\Theta} g_z(-\infty) \ ,
  \\
  \langle \gp_+(\infty) \rangle &=&
  G_{+,+}(\infty,0) \langle \gp_z(\tx) \rangle
  \nonumber \\
  &=& - \sqrt2 a^* b^* e^{-[\vartheta_+(\infty)-\vartheta_+(0)]}
  \nonumber \\ && \times
  e^{-\Theta/2} g_z(-\infty) \ .
\end{eqnarray}
\end{subequations}
These results were obtained already in Ref.~\cite{PS2}. In terms
of $\mathbf{g} $ they mean
\begin{subequations}
\begin{eqnarray}
  \langle g_z(\infty) \rangle &=&
  (|a|^2-|b|^2) e^{-\Theta} g_z(-\infty) \ ,
  \label{gz.final}
  \\
  \langle g_\perp(\infty) \rangle &=&
  \sqrt2 |a||b| e^{-3\Theta/4} g_z(-\infty) \ .
\end{eqnarray}
\end{subequations}
We recall that the averages $\langle g_x(\infty) \rangle = \langle
g_y(\infty) \rangle=0$ vanish due to the presence of the
longitudinal noise component.  The correlations follow from
Eqs.~(\ref{C.prop}) with $t=\infty $ and ${t_{0}}=\tx$:
\begin{subequations}
\label{Cp.final}
\begin{eqnarray}
  \Cp_{zz}(\infty)
  &=& \frac 13 \left[ 1+2(1-6|a|^2|b|^2)e^{-3 \Theta}\right]
  g_z^2(-\infty) \ ,
  \\
  \Cp_{+-}(\infty)
  &=& \frac 13 \left[ 1-(1-6|a|^2|b|^2)e^{-3 \Theta}\right]
  g_z^2(-\infty) \ .
\end{eqnarray}
\end{subequations}
In the presence of regular transitions, the transition probabilities
and their fluctuations -- given in Eqs.~(\ref{prob.nonoise}) in the
absence of $\Delta$ -- finally become
\begin{subequations}
\label{prob.final}
\begin{eqnarray}
  \langle \Pi _{\uparrow \uparrow /\uparrow \downarrow }(\infty )\rangle
  &=&
  \frac{1}{2}[1\pm (|a|^{2}-|b|^{2})e^{-\Theta }]
\end{eqnarray}
and
\begin{eqnarray}
  \langle [\delta \Pi _{\uparrow \uparrow /\uparrow \downarrow }(\infty
  )]^{2}\rangle &=&\frac{1}{12}[1-3(1-4|a|^{2}|b|^{2})e^{-2\Theta }  \nonumber
  \\
  &&+2(1-6|a|^{2}|b|^{2})e^{-3\Theta }] \ .
\end{eqnarray}
\end{subequations}
To obtain the explicit dependence on the LZ parameter one may use
$|a|^2 = e^{-2\pi \gamma^2}$ and the unitarity relation
$|b|^2=1-|a|^2$.

\section{Response spectrum and intensity}
\label{sec.spect}

Let a short pulse of an effective field directed along axis $i$
act on the LZ system at some moment of time $t$. Its response at
a later moment of $t^{\prime}=t+\tau$ is determined by the
correlation function, $K_{i,j}(t,\tau)=\langle
s_{i}(t)s_{j}(t+\tau )\rangle $. It is reasonable to
measure the spectral content of the response at a fixed time of
initial pulse $t$. It is given by the Fourier-transform
\begin{equation}\label{spectrum}
  K_{i,j}(t;\omega)=\int_0^{\infty}
  K_{i,j}(t,\tau)e^{i\omega\tau}d\tau \ .
\end{equation}
According to general relations (\ref{two-time}) and (\ref{gen}), in
the limit $\tau \rightarrow \infty $ such a function vanishes or
saturates to a finite value.  For example, in the case $i=j=z$ and $t
\gg \tx$ it is equal to $\tilde{C}_{zz}(t) \textrm{exp} \left(
  -\int\nolimits_{t}^{\infty } F_\perp (\nu \tau ^{\prime })d\tau
  ^{\prime }\right) $.  Therefore, the spectral density of the
response contains a $1/\omega$ component at small frequency with an
intensity that depends on the time of excitation. This is not
surprising since the LZ process violates the homogeneity of time. At
infinite time $t$ this intensity reaches a finite limit
$\tilde{C}_{zz}(\infty )$. It happens since the noise becomes
ineffective at sufficiently large time and does not randomize the
quantum amplitudes any more.  Therefore, any quantum amplitude or a
component of the Bloch vector $\mathbf{g}$ changes after this time in
an almost deterministic way. On the other hand, the spectral intensity
remains finite even at $t\rightarrow \infty $ since the value of
$\mathbf{g}$ reached at the moment, when the noise becomes
ineffective, is random.

Another (incoherent) contribution to the spectral density arises from
the variation of the correlation function at a finite time:
\begin{eqnarray}
  K_{zz}(\tau ,t)-K_{zz}(\infty ,t)=K_{zz}(\infty ,t)
  \nonumber \\ \times
  \left[ \exp \left(
      \int_{t+|\tau |}^{\infty } F_\perp (\nu \tau ^{\prime })
      d\tau ^{\prime}\right) -1\right]   \ .
  \label{remnant}
\end{eqnarray}

To estimate the spectral width of this contribution we assume that
either time $t$ is large enough or the noise is weak. Then the
exponent in equation (\ref{remnant}) can be expanded to the linear
term. Its spectral width is easily estimated as $\Delta \omega
_{1}=1/\tau _{\mathrm{acc}}$. The spectral intensity of this line at
$t\thicksim t_{\mathrm{acc}}$ can be estimated as $I_{1}\varpropto
\tilde{C}_{zz}(\infty )\left\langle \eta _{x}^{2}+\eta
  _{y}^{2}\right\rangle /\nu $. It decreases rapidly with growing time
$t$.

If the LZ processes are repeated randomly with the average time
$\tau_{\mathrm{coll}}$ between the events (as it happens in a gas of
colliding atoms or molecules), the $1/\omega$-line is smeared out to
the width $\Delta \omega _{0}\thicksim 1/\tau _{\mathrm{coll}}\ll
\Delta \omega _{1}$, whereas its average intensity $I_{0}$ is
proportional to $\tilde{C}_{zz}(\infty )$.  The width of the
incoherent line remains the same $\Delta \omega _{1}=1/\tau
_{\mathrm{acc}}$, but its intensity is stabilized at a value
$I_{1}(\tau_{\mathrm{coll}})\varpropto I_{1}\varphi
(\tau_{\mathrm{coll}}/\tau _{\mathrm{acc}})$ the smaller the larger is
$\tau_{\mathrm{coll}}$ [here $\varphi (x)$ is a function describing
the decay of the noise correlations, which depends on details].

Thus, the main contribution to the spectral width of the
population or induced field fluctuations consists of two narrow
lines. The first has the width determined by collisions and a
permanent intensity per particle. The second has a permanent
width, but its intensity is determined by collision time. In the
rarefied gas, or more generally if $\tau _{\mathrm{coll}}\gg \tau
_{\mathrm{acc}}$, the first line is narrower and stronger than the
second one.

\section{Conclusions}
\label{sec.conc}

We have calculated the correlation functions of the Bloch vector
components in the LZ process subject to a fast Gaussian noise.  These
correlators determine the linear response of the LZ system to a weak,
time dependent probe signal. Two-time correlation functions are
factorizable at the time scale much larger than the noise correlation
time $\tau _{\mathrm{n}}$ due to statistical independence of random
processes. Thus the main problem is the calculation of simultaneous
averages.  The condition of the fast noise allows to use for these
averages the ladder graphs only. The resulting Bethe-Salpeter equation
can be reduced to differential equations which are exactly solvable.
In general, fluctuations are strong, of the same order of magnitude as
the average values.

Since the noise is ineffective for transitions at sufficiently large
time $ t\gg \tau _{\mathrm{acc}}$, the transition probabilities or the
components of the Bloch vector remain deterministically coherent after
this time.  Therefore, the spectrum of the fluctuations contains a
narrow $1/\omega$ line, whose intensity is determined by the
distribution of these values at the time when the noise became
ineffective. Besides of this line, there exists another narrow line
with a width of about $1/\tau _{\mathrm{acc}}$ and an intensity
depending on the pulse time and going to zero when this time goes to
infinity.

\begin{acknowledgments}
  V.L.P. acknowledges support by DOE under the grant
  DE-FG03-96ER45598, NSF under the grants DMR-0321572 and DMR 0103455,
  Texas A{\&}M University Telecommunications and Informatics Task
  Force, the Humboldt Foundation and the Institute of Theoretical
  Physics, University of Cologne for the hospitality.  S.S.
  acknowledges support by Deutsche Forschungsgemeinschaft under the
  project SFB 608.
\end{acknowledgments}

%\newpage
\appendix

\section{Details for Correlators}

Here we provide some details on the calculation of correlation
propagators.

\subsection{Propagator $G_{+-,zz}$}

\begin{figure}[htbp]
\centering
\epsfig{file=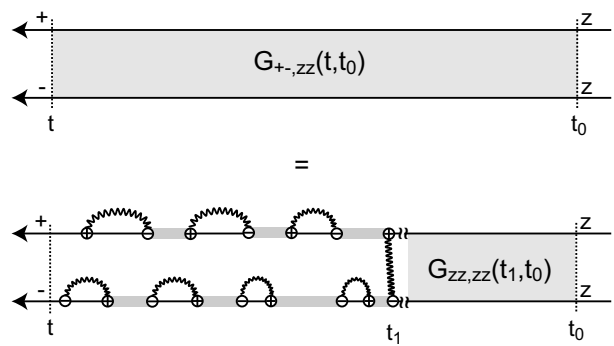,width=\figwidth}
\caption{ The diagrams of $G_{+-,zz}$ are analogous to the diagrams of $%
G_{zz,+-}$ with the only difference that now the excess vertices are the
last vertices.}
\label{figpmzz}
\end{figure}

The calculation of the propagator $G_{+-,zz}$ can be performed largely
in parallel to the calculation of $G_{zz,+-}$ in
Sec.~\ref{sec.G.zzpm}.  Starting point is the equation
\begin{eqnarray*}
G_{+-,zz}(t,t_{0}) &=&\int_{t_{0}}^{t}dt^{\prime }\int_{t_{0}}^{t}d\bar{t}
^{\prime }\int_{-\infty }^{\infty }d\bar{t}^{\prime \prime }\int_{-\infty
}^{\infty }dt^{\prime \prime }  \nonumber \\
&&\times \langle \tilde{\eta}_{-}(t^{\prime })\tilde{\eta}_{+}(\bar{t}
^{\prime })W_{t_{0}}^{t^{\prime }}(t^{\prime \prime })W_{t_{0}}^{\bar{t}
^{\prime }}(\bar{t}^{\prime \prime })\rangle \ .
\end{eqnarray*}
Using the Wick theorem, we perform the contractions
\begin{eqnarray*}
G_{+-,zz}(t,t_{0}) &=&\int_{t_{0}}^{t}dt_{1}G_{+,+}(t,t_{1})G_{-,-}(t,t_{1})
\nonumber \\
&&\times f(\nu t_{1})G_{zz,zz}(t_{1},{t_{0}}) \ ,
\end{eqnarray*}
where $t_{1}$ is the time of the last rung of the ladder (last means
closest to $t$, see Fig.~\ref{figpmzz}). After a substitution from
$t_{1}$ to $y_{1}$ the integral is elementary with the result
\[
G_{+-,zz}(t,{t_{0}})=\frac{1}{3}\left( 1-e^{-3[\vartheta _{z}(t)-\vartheta
_{z}({t_{0}})]}\right)  \ .
\]

\subsection{Propagator $G_{+-,+-}$}

\begin{figure}[htbp]
\centering
\epsfig{file=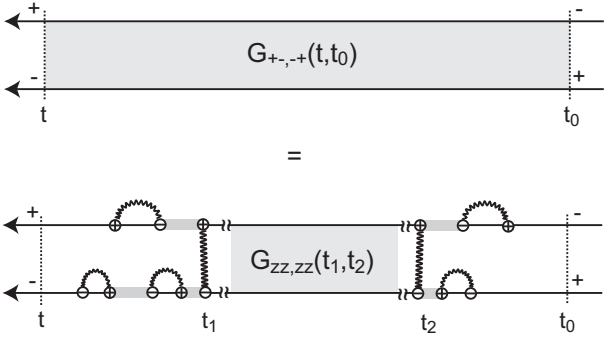,width=\figwidth}
\caption{ The diagrams of $G_{+-,-+}$ as similar to these of $G_{+-,+-}$
with the difference that the polarity of the first excess vertices is
inverted. Therefore, no disconnected diagrams can be formed.}
\label{figpmmp}
\end{figure}

From the equations of motion (\ref{iterated}) one extracts
\begin{widetext}
\begin{eqnarray*}
  G_{+-,+-}(t,\ti) &=&
  1 - \int_\ti^t dt' \int_{-\infty}^\infty dt'' \int_\ti^{t''} dt'''
  \langle \etap_-(t') W_\ti^{t'}(t'') \etap_+(t''') \rangle
  \nonumber \\ &&
  - \int_\ti^t d\bar t' \int_{-\infty}^\infty d \bar t'' \int_\ti^{\bar t''}
  d \bar t'''
  \langle \etap_+(\bar t') W_\ti^{\bar t'}(\bar t'') \etap_-(\bar t''') \rangle
  \nonumber \\ &&
  +  \int_\ti^t dt' \int_{-\infty}^\infty dt'' \int_\ti^{t''} dt'''
  \int_\ti^t d\bar t' \int_{-\infty}^\infty d \bar t'' \int_\ti^{\bar t''}
  d \bar t'''
%  \nonumber \\ && %%%
  \langle \etap_-(t') W_\ti^{t'}(t'') \etap_+(t''')
  \etap_+(\bar t') W_\ti^{\bar t'}(\bar t'') \etap_-(\bar t''') \rangle \ .
\end{eqnarray*}
After averaging, the sum over the ladder diagrams can be captured by a
Bethe-Salpeter equation
\begin{eqnarray*}
  G_{+-,+-}(t,\ti) &=& G_{+,+}(t,\ti) G_{-,-}(t,\ti)
  \nonumber \\ &&
  +  \int_\ti^t dt_1 \int_\ti^{t_1} dt_2 G_{+,+}(t,t_1) G_{-,-}(t,t_1)
  f(\nu t_1) G_{zz,zz}(t_1,t_2) f(\nu t_2) G_{+,+}(t_2,\ti) G_{-,-}(t_2,\ti)
\end{eqnarray*}
where $t_1$ is the time of the last rung and $t_2$ the time of the
first rung (see Fig. \ref{figpmpm}).  All contractions in between sum
up to $G_{zz,zz}(t_1,t_2)$.  Since the latter propagator is explicitly
known, the integrals can be performed directly, yielding
Eq.~(\ref{G.+-.+-})

\begin{figure}[htbp]
\centering
\epsfig{file=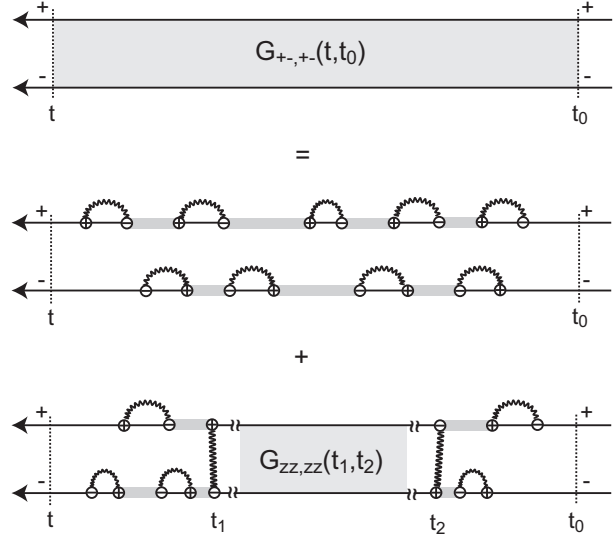,width=\figwidth}
\caption{$G_{+-,+-}$ consists of diagrams with excess vertices as first and
last vertices on both time legs. There are disconnected diagrams (where the
``ladder'' has no rungs) and connected diagrams (where the ladder has
rungs). In the second case, the excess vertices are are connected by rungs
as $t_1$ and $t_2$ and between $t_1$ and $t_2$ additional pairs can be
contracted as in $G_{zz,zz}(t_1,t_2)$.}
\label{figpmpm}
\end{figure}

\subsection{Propagator $G_{+-,-+}$}
Similar to the previous propagator, one now has
\begin{eqnarray*}
  G_{+-,-+}(t,\ti) &=&
  \int_\ti^t dt' \int_{-\infty}^\infty dt'' \int_\ti^{t''} dt'''
  \int_\ti^t d\bar t' \int_{-\infty}^\infty d \bar t'' \int_\ti^{\bar t''}
  d \bar t'''
%  \nonumber \\ && %%%
  \langle \etap_-(t') W_\ti^{t'}(t'') \etap_-(t''')
  \etap_+(\bar t') W_\ti^{\bar t'}(\bar t'') \etap_+(\bar t''') \rangle \ .
\end{eqnarray*}
The configuration of the excess vertices necessitates at least two
rungs at times $t_1$ and $t_2$ (see. Fig.~\ref{figpmmp}).  The
contractions in between again sum up to $G_{zz,zz}(t_1,t_2)$, leading to
\begin{eqnarray*}
  G_{+-,-+}(t,\ti) &=&
  \int_\ti^t dt_1 \int_\ti^{t_1} dt_2 G_{+,+}(t,t_1) G_{-,-}(t,t_1)
  f(\nu t_1) G_{zz,zz}(t_1,t_2) f(\nu t_2) G_{+,+}(t_2,\ti) G_{-,-}(t_2,\ti)
\end{eqnarray*}
\end{widetext}
and eventually to Eq. (\ref{G.+-.-+}).

\begin{thebibliography}{9}
\bibitem{note1} {%
    The evolution matrix connects the values of the ``slow''
    amplitudes $C_{\uparrow',\downarrow}'(t)$ which are defined by
    representing the state vector in a form $|t\rangle
    =C_{\uparrow}'(t) \ e^{-i \Phi_{\uparrow}(t)} |{\uparrow}\rangle
    +C_{\downarrow}'(t) \ e^{-i \Phi_{\downarrow}(t)}
    |{\downarrow}\rangle $.  The dynamic phases
    $\Phi_{\uparrow,\downarrow}(t)= (1/\hbar) \int_0^t dt'
    E_{\uparrow,\downarrow}'(t')$ depend on the diabatic eigenstates
    of the Hamiltonian (\ref{H.LZ}).}
\end{thebibliography}
\end{document}